# Correspondence Analysis between the Location and the Leading Causes of Death in the United States

**Rena Sandy H. Baculinao**[1] and **Roel F. Ceballos**[2]

[1]Local Government Unit, Municipality of Carmen
Davao Del Norte, Philippines
Email: renasandybaculinao12@gmail.com

[2]Department of Mathematics and Statistics
University of Southeastern Philippines, Bo. Obrero, Davao City
Email: roel.ceballos@usep.edu.ph

**ABSTRACT**

*Correspondence Analysis analyzes two-way or even multi-way tables with each row and column becoming a point on a multidimensional graphical map called biplot. It can be used to extract the essential dimensions allowing simplification of the data matrix. This study aims to measure the association between the location and the leading causes of death in the United States and to determine the location where a particular leading cause of death is associated. The research data consists of two variables (location and leading causes of death in the United States of America) with 510 data points. Results show that there exists a significant association between the location and the leading causes of death in the United States, and 61.6% of the variance in the model was explained by Dimensions 1 and 2. Furthermore, Connecticut, Massachusetts, New Hampshire, New Mexico, Ohio, Rhode Island, West Virginia, and Wisconsin are associated with unintentional injuries. In contrast, Arizona, California, Georgia, Minnesota, North Dakota, South Carolina, South Dakota, Texas, Vermont, and Washington are associated with Alzheimer's disease. Cancer is highly associated with Delaware and Maryland. CLRD is associated with Alabama, Arkansas, Indiana, Iowa, Kansas, Kentucky, Maine, Nebraska, and Oklahoma. In the District of Columbia, Michigan, New Jersey, New York, and Pennsylvania, heart disease is the leading cause of death. Influenza and pneumonia are associated with Hawaii, Mississippi, and Tennessee. Also, kidney disease is associated with Illinois, Louisiana, Missouri, North Carolina, and Virginia. Florida is associated with stroke. Lastly, suicide is associated with Alaska, Idaho, Montana, Nevada, Oregon, Utah, and Wyoming. These findings may serve as additional insights in understanding the complexity of human diseases and mortality. The United States government and its health organizations should create relevant programs and policies to address the leading cause of death in different states.*

## 1. INTRODUCTION

Mortality statistics are usually presented using the top leading causes of death in a country or a state. A report of this kind first appeared in 1949, where an official tabulation of death by cause was made available to the public. Since then, the data on the leading causes of death report has been generated continuously and used as a basis to plan and manage public health. Understanding the association of the leading causes of death and the location in which they are prevalent will help the state to appropriately focus their efforts and resources on their top priority health problems.

Several studies have explored the use of multivariate methods to analyze health data. Carillo, Largo, and Ceballos (2017) analyzed the Philippine health data using principal component analysis. Ploeg et al. (1993) studied the risk factors for primary dysfunction after a liver transplant using multivariate analysis. This paper aims to know which among the leading causes of death is associated with the different location or state in the United States using Correspondence Analysis. Results and findings of the study may be used by the government of the United States and the health organizations to create plans or programs in preventing and managing the leading causes of death in a specific state where a cause of death is highly associated.

## 2. METHODS

### 2.1 Sources of Data

The data was taken from catalog.data.gov website consisting of age-adjusted death rates for the ten leading causes of death in the United States. The number of deaths recorded in the year 2016 was utilized for the analysis. The data consist of two variables: the location and the leading causes of death in the United States for the year 2016. The leading causes of death are unintentional injuries, Alzheimer's disease, cancer, chronic lower respiratory disease, diabetes, heart disease, influenza and pneumonia, kidney disease, stroke, and suicide (Appendix A). Moreover, the location is the 50 states of the United States and the District of Columbia.

### 2.2 Mathematical Concept of Correspondence Analysis

Correspondence Analysis is a multivariate method used as an exploratory data technique on categorical data for which no specific hypotheses have been formed (Storti, 2010). The following are the step by step computational method in performing the analysis (Rencher, 2002):

1. Established an $n \times m$ data matrix, $K$, where n is the number of rows, and m is the number of columns. The elements of $matrix\ K$ must be non-negative and that no row total and column total is equal to zero.
2. Compute $P$, the proportion matrix, by dividing the elements of $K$ by the total of all numbers in $K$. Mathematically, proportion matrix ($P$) is equal to
$P = \{p_{ij}\} = \left\{\frac{k_{ij}}{k}\right\}$.
3. Compute the total of the rows of matrix $P$ and columns of matrix $P$, putting the results in vectors $r$ and $c$. That is, using standard matrix notation,
$$r = P1$$
$$c = P'1$$
where **1** is a vector of ones.
4. Take the square roots of the vectors $r$ and $c$ and transform them into diagonal matrices and then take the inverse of the resulting square matrices.
$$D_r = [diag(r)]^{\frac{-1}{2}}$$
$$D_c = [diag(c)]^{\frac{-1}{2}}$$
5. Compute the scaled matrix, $A$ using the following matrix formula.
$$A = D_r P D_c$$
6. Calculate $B$, $W$, and $C$ by solving the Singular Value Decomposition (SVD) of $A$.
$$\langle B, W, C \rangle = SVD(A)$$
7. Compute the coordinate matrices, $F$ and $G$, as follows:
$$F = D_r BW$$
$$G = D_c CW'$$
8. Compute the eigenvalues, $V$.
$$V = WW''$$
9. Compute the row distances ($d_i$) and column distances ($d_j$).

$$d_i = \sum_j \left(\frac{1}{p_{.j}}\right)\left(\frac{p_{ij}}{p_{i.}} - p_{.j}\right)^2$$

$$d_j = \sum_j \left(\frac{1}{p_{i.}}\right)\left(\frac{p_{ij}}{p_{.j}} - p_{i.}\right)^2$$

10. The weighs, $w_i$, and $w_j$, come from the vectors **r**, and **c** were formed in step 3.

$$w_i = \{r_i\} \text{ and } w_j = \{c_j\}.$$

Additionally, the Signed Chi-Square statistic is computed to have a more objective basis in determining the association between the location and causes of death in the United States. To compute for the Signed Chi-Square statistic, we first compute the chi-square statistic using the formula

$$X^2 = \sum_{i=1}^{a}\sum_{j=1}^{b} \frac{\left(n_{ij} - \frac{n_{i+}n_{j+}}{n}\right)^2}{\frac{n_{i+}n_{j+}}{n}}$$

where $n_{ij} = n_{i+} = \sum_{j=1}^{b} n_{ij}$ and $n_{j+} = \sum_{i=1}^{a} n_{ij}$ are the marginal frequencies, and $n$ is the total number of individuals in the sample. The last step is to assign a negative (if the expected value is greater than the observed value) or positive sign (if the observed value is greater than the expected value) to the computed chi-square value.

## 3. RESULTS AND DISCUSSION

In Table 1, the Chi-Square statistic value is 28,696.584, with a p-value of <0.01 is presented. Therefore, it is concluded that there exists a significant association between the leading causes of death and the location in the United States, as analyzed in eight (8) dimensions. The singular value column gives the square roots of the eigenvalues, which describes the maximum canonical correlation between the categories of the variables in the analysis for any given dimension. The values in the proportion of inertia give the percentage of variance that each dimension explains in the model. That is, Dimensions 1 and 2 explain 61.6% of the variances in the model.

*Table 1. Summary of the Main Statistical Results*

| Dimension | Singular Value | Inertia | Proportion of Inertia | |
|---|---|---|---|---|
| | | | Accounted for | Cumulative |
| 1 | 0.078 | 0.006 | 0.429 | 0.429 |
| 2 | 0.051 | 0.003 | 0.187 | 0.616 |
| 3 | 0.038 | 0.001 | 0.104 | 0.719 |
| 4 | 0.035 | 0.001 | 0.086 | 0.805 |
| 5 | 0.029 | 0.001 | 0.060 | 0.865 |
| 6 | 0.026 | 0.001 | 0.049 | 0.914 |
| 7 | 0.024 | 0.001 | 0.040 | 0.954 |
| 8 | 0.019 | 0.000 | 0.025 | 0.979 |

*Chi-square = 28696.584 ; p-value=<0.001*

The results of the signed chi-square and the biplot are presented in Appendix B and C, respectively. In this case, the biplot does not help determine which location is associated with a particular leading cause of death. Thus, we interpret the association based on the signed chi-square. The negative values indicate less association or lower similarity, and positive values indicate higher association and similarity. For every row/location, the highest positive signed chi-square value per row or location was marked(yellow). We traced those cells with marked(yellow) and considered their corresponding row(location) and column(leading cause) to be associated. In every row/location, cells with the lowest

International Journal of Ecological Economics and Statisticswere highlighted with red. These locations are less associated with the column variable (the leading cause of death).

Results in Table 2 indicate that Connecticut, Massachusetts, New Hampshire, New Mexico, Ohio, Rhode Island, West Virginia, and Wisconsin are associated with unintentional injuries. Alzheimer's disease is the most commonly associated leading cause of death with Arizona, California, Georgia, Minnesota, North Dakota, South Carolina, South Dakota, Texas, Vermont, and Washington. Maryland and Delaware should focus their health programs on Cancer. The CLRD is associated with Alabama, Arkansas, Indiana, Iowa, Kansas, Kentucky, Maine, Nebraska, and Oklahoma. Health policies that will mitigate death due to heart disease should be made in Michigan, District of Columbia, New Jersey, New York, and Pennsylvania. Influenza and pneumonia are associated with Hawaii, Mississippi, and Tennessee. Excellent facilities for kidney treatment and management should be present in Illinois, Louisiana, Missouri, North Carolina, and Virginia. Stroke is the top health issue in Florida, while suicide is associated with Alaska, Idaho, Montana, Nevada, Oregon, Utah, and Wyoming.

*Table 2. Highly Associated Locations of Leading Causes of Death*

| Leading Cause of Death | Highly Associated Locations |
|---|---|
| Unintentional injuries | Connecticut, Massachusetts, New Hampshire, New Mexico, Ohio, Rhode Island, West Virginia, Wisconsin |
| Alzheimer's disease | Arizona, California, Connecticut, Georgia, Minnesota, North Dakota, South Carolina, South Dakota, Texas, Vermont, Washington |
| Cancer | Delaware, Maryland, |
| CLRD | Alabama, Arkansas, Indiana, Iowa, Kansas, Kentucky, Maine, Nebraska, Oklahoma, Tennessee, |
| Heart disease | District of Columbia, Michigan, New Jersey, New York, Pennsylvania, |
| Influenza and pneumonia | Hawaii, Mississippi, |
| Kidney disease | Illinois, Louisiana, Missouri, North Carolina, Virginia, |
| Stroke | Florida, |
| Suicide | Alaska, Colorado, Idaho, Montana, Nevada |

Results in Table 3 reveal that unintentional injury is less associated with Arkansas, California, Illinois, Michigan, and Nebraska. Cancer is not a top priority for health programs in Alabam, Louisiana, Mississippi, North Dakota, Oklahoma, Tennessee, and Utah. Also, CLRD is not common in the District of Columbia, Hawaii, and New Jersey. We also found out that Alaska, Colorado, Georgia, Maine, Minnesota, North Carolina, Oregon, South Carolina, Virginia, West Virginia, Wisconsin should not focus their efforts on Heart Disease. Diabetes is the least concern of Delaware and Missouri, while for Florida, Indiana, and Texas, their least concern is Influenza and pneumonia.

*Table 3. Less Associated Locations of Leading Causes of Death*

| Leading Cause of Death | Less Associated Locations |
|---|---|
| Unintentional injuries | Arkansas, California, Illinois, Michigan, Nebraska, |
| Alzheimer's disease | Kansas, Maryland, Massachusetts, Montana, Nevada, New Mexico, New York, Pennsylvania, Wyoming |
| Cancer | Alabama, Louisiana, Mississippi, North Dakota, Oklahoma, Tennessee, Utah |
| CLRD | District of Columbia, Hawaii, New Jersey, |
| Diabetes | Delaware, Missouri, |
| Heart disease | Alaska, Colorado, Georgia, Maine, Minnesota, North Carolina, Oregon, South Carolina, Virginia, West Virginia, Wisconsin |
| Influenza and pneumonia | Florida, Indiana, Texas |
| Kidney disease | Arizona, Idaho, Iowa, South Dakota, Vermont, Washington |
| Stroke | Connecticut, Kentucky, New Hampshire, Rhode Island, |
| Suicide | Ohio |

50

Based on the results found in Tables 2 and 3, the United States government and the health organizations should create plans and programs with regards to the specific leading cause of death associated with every location. Like in the states of Alaska, Idaho, Montana, Nevada, Oregon, Utah, and Wyoming, the United States government should strengthen mental health programs to prevent or reduce suicide rates.

## Appendix A: Definition of the Leading Causes of Death

- **Unintentional injuries** – or also known as accidents, ranked as the 4th leading cause of death in the United States. Though accidents are unintentional, there are many ways to prevent accidental death and injury from happening. Focusing on road safety measures such as using the seat belt and raising awareness of the dangers of driving while drunk is some ways of preventing accidents.
- **Alzheimer's disease** – is the damage and death of neurons that eventually impair the capacity to carry out bodily functions such as walking and swallowing. The following are some common signs and symptoms of Alzheimer's disease: memory loss, forgetting where things are placed and losing the ability to retrace steps, changes in mood and personality, and have a problem in speaking or writing words. Though it is hard to prevent the disease, some steps may help to delay Alzheimer's disease from happening, such as having regular exercise, eating a healthy balanced diet, stopping from smoking cigarettes, and drinking alcoholic drinks.

- **Cancer** – is the uncontrolled growth and spread of abnormal cells that can interfere with essential life-sustaining systems and may result in death. Some diseases are preventable like cancers caused by smoking a cigarette and drinking alcoholic drinks. Cancers are preventable for all cases related to being overweight, obese, inactive, and poor nutrition.

- **Chronic lower respiratory diseases (CLRD)** – is a collection of lung diseases caused by airflow blockage and breathing-related issues, including primarily chronic obstructive pulmonary disease, bronchitis, emphysema, and asthma. Signs and symptoms include breathlessness, cough with phlegm, and chest infections. These are some ways to prevent CLRD: quit smoking, avoid second-hand smoke, avoid air pollution and dust, and avoid chemical fumes.

- **Diabetes** – is a disease in which the body is no longer able to control blood glucose that leads to high levels of blood glucose in the body. Some signs and symptoms of diabetes are frequent urination, excessive thirst, extreme hunger, sudden vision changes, and dehydrated skin.

- **Heart disease** – is also the leading cause of death globally. Heart disease is the disorder of the heart. To prevent deaths from heart disease is to protect the heart and know the warning signs and symptoms of a heart attack. The following are signs and symptoms of heart diseases: chest pain, discomfort in the upper body (arms, jaw, neck, or upper stomach), breathlessness, nausea, lightheadedness, or cold sweats.

- **Influenza and pneumonia** – Influenza is a contagious viral infection and one of the severe illnesses in the winter season since the virus survives and is transmitted better in cold temperatures. Influenza is transferred from one person to another when an infected person coughs or sneezes. Influenza can complicate to pneumonia, a condition that can cause inflammation of the lungs. These are some signs and symptoms of influenza: fever, headache, cough, nasal congestion, and loss of appetite. While for pneumonia, signs, and symptoms include: fever, cough, rapid breathing, chest pains, and loss of appetite.

- **Kidney disease** – is a condition in which kidneys are damaged and can no longer filter blood. If the kidneys can no longer function, waste from the blood remains in the body and may cause other health problems. Some signs and symptoms of kidney disease include general ill feeling and fatigue, appetite loss, nausea, and weight loss without trying to lose weight. The following are some symptoms when kidney disease becomes severe: bone pain, numbness or swelling in the hands and feet, breath odor, menstrual period stops, and frequent hiccups. This disease is prevented by avoiding excessive intake of alcohol, maintaining a healthy weight, and management with a healthcare professional.

- **Stroke** – is a condition that develops as a result of problems with the blood vessels that supply the brain. Some signs and symptoms of stroke include numbness or weakness in some parts of the body, trouble in speaking or difficulty in understanding speech, and discomfort in seeing in one or both eyes. These are some steps to prevent stroke and its complications: maintaining a healthy weight, getting enough exercise, not smoking and limiting alcohol use, managing cholesterol levels, and controlling blood pressure.

- **Suicide** – is a significant health problem and the 10th leading cause of death in the United States. The following are some signs of a suicidal person: sleeping too little or too much, acting anxious or agitated, and withdrawing or feeling isolated.

## APPENDIX B Signed Chi-Square Tables

| States | Causes | | | | | | | | | |
|---|---|---|---|---|---|---|---|---|---|---|
| | Unintentional injuries | Alzheimer's disease | Cancer | CLRD | Diabetes | Heart disease | Influenza and pneumonia | Kidney disease | Stroke | Suicide |
| Alabama | -33.55 | 38.99 | -84.46 | 48.73 | -77.15 | 43.07 | 0.02 | 3.52 | 24.45 | -5.58 |
| Alaska | 128.66 | -29.59 | 1.86 | -0.28 | -0.10 | -32.43 | -5.43 | -8.97 | -4.10 | 205.56 |
| Arizona | 139.78 | 197.49 | -16.95 | 117.41 | 84.72 | 99.86 | -29.75 | -287.88 | -48.06 | 127.39 |
| Arkansas | -57.18 | 4.77 | -44.56 | 53.00 | -1.74 | 28.45 | 0.03 | 24.52 | -2.32 | 0.44 |
| California | -501.21 | 1401.76 | 0.02 | -180.3 | 169.57 | -40.81 | 142.78 | -359.28 | 168.53 | -7.08 |
| Colorado | 220.48 | 43.78 | -3.67 | 110.49 | -19.76 | -204.82 | -39.03 | -51.51 | 0.00 | 512.98 |
| Connecticut | 38.41 | -33.33 | 15.90 | -30.33 | -28.05 | 11.29 | 0.91 | 2.47 | -40.19 | -14.20 |
| Delaware | -0.27 | -6.81 | 14.32 | 2.89 | -13.26 | -5.24 | -7.95 | 11.97 | 3.61 | -5.37 |
| District of Columbia | 43.17 | -37.30 | -0.73 | -56.48 | -1.90 | 49.06 | -2.24 | -11.92 | -0.03 | -20.45 |
| Florida | 53.37 | -203.31 | 9.78 | 42.94 | -0.54 | -9.65 | -240.35 | -70.07 | 213.6 | -5.65 |
| Georgia | -1.02 | 100.90 | -13.94 | 12.03 | -7.01 | -21.67 | -6.65 | 99.89 | 5.12 | 4.77 |
| Hawaii | -6.84 | -0.69 | 0.12 | -122.0 | -3.65 | -0.78 | 442.59 | 1.20 | 20.72 | -0.15 |
| Idaho | 4.15 | 2.46 | -0.85 | 14.98 | -0.80 | -6.98 | -8.31 | -12.55 | -0.34 | 77.10 |
| Illinois | -91.44 | -88.16 | 59.15 | -22.78 | -34.41 | 4.99 | 15.70 | 170.23 | 3.54 | -62.63 |
| Indiana | -7.83 | 0.63 | -1.88 | 140.6 | 16.57 | -15.79 | -37.18 | 66.17 | -11.02 | 0.16 |
| Iowa | -10.10 | 21.14 | -0.05 | 68.73 | -0.32 | 1.06 | -4.84 | -65.33 | -13.74 | -2.38 |
| Kansas | -1.36 | -43.65 | -0.21 | 36.32 | -0.18 | -6.11 | 4.55 | 10.99 | 2.02 | 23.68 |
| Kentucky | 50.55 | -43.97 | -0.54 | 230.0 | 4.77 | -29.20 | -0.15 | 29.48 | -72.46 | -1.06 |
| Louisiana | 1.32 | 43.43 | -46.36 | -39.95 | 0.06 | 24.70 | -18.93 | 133.19 | -0.07 | -5.14 |
| Maine | 8.24 | -0.52 | 14.68 | 23.43 | 6.84 | -36.92 | -4.11 | -1.17 | -5.80 | -0.08 |
| Maryland | -75.02 | -311.41 | 66.62 | -110.4 | 0.04 | 41.98 | 27.76 | -0.55 | 40.21 | -39.32 |
| Massachusetts | 150.97 | -135.14 | 98.71 | -37.56 | -54.28 | -16.27 | 61.03 | 27.77 | -32.47 | -68.31 |
| Michigan | -48.49 | 0.45 | -28.99 | -0.14 | -17.91 | 228.55 | -20.33 | -0.26 | -14.66 | -43.03 |
| Minnesota | 35.95 | 138.90 | 101.63 | 1.94 | 4.45 | -282.57 | -74.09 | -33.18 | 3.05 | 8.49 |
| Mississippi | -10.82 | 4.96 | -58.85 | 33.39 | 14.28 | 5.00 | 42.61 | 41.75 | -0.07 | -47.07 |
| Missouri | 0.06 | -33.59 | -34.86 | 73.00 | -44.72 | 9.50 | -0.01 | 117.95 | -3.87 | 16.05 |
| Montana | 6.07 | -21.72 | -2.43 | 58.63 | 3.10 | -4.06 | -6.45 | -2.49 | -7.13 | 75.06 |
| Nebraska | -19.77 | -0.48 | 4.29 | 71.57 | 5.93 | -16.68 | 9.60 | -13.21 | -0.15 | -0.16 |
| Nevada | -5.26 | -136.15 | -14.57 | 89.69 | -33.02 | 65.38 | 18.37 | -75.21 | -33.92 | 135.55 |
| New Hampshire | 64.04 | -12.93 | 22.11 | 0.01 | -3.79 | -9.92 | -2.53 | -16.01 | -29.54 | 10.68 |
| New Jersey | -33.03 | -116.97 | 37.76 | -233.7 | -9.46 | 245.18 | -13.98 | 40.06 | -25.74 | -201.77 |
| New Mexico | 181.42 | -39.49 | -28.47 | 15.44 | 47.22 | -27.17 | 0.92 | -3.30 | -1.75 | 108.62 |
| New York | -367.94 | -1611.90 | 49.52 | -430.5 | -59.92 | 1718.6 | 847.09 | -76.18 | -417.9 | -304.08 |
| North Carolina | 12.93 | 42.65 | 1.91 | 19.71 | 19.25 | -250.54 | 31.79 | 91.30 | 25.96 | -4.45 |
| North Dakota | 0.56 | 42.29 | -3.68 | -2.28 | -0.21 | -3.20 | 3.51 | -0.20 | -0.04 | 16.53 |

| | | | | | | | | | |
|---|---|---|---|---|---|---|---|---|---|
| Ohio | 132.12 | -0.18 | -12.10 | 11.27 | 1.74 | -2.90 | -1.59 | 2.96 | -7.08 | -32.70 |
| Oklahoma | 12.31 | -13.05 | -80.81 | 97.83 | 30.94 | 49.23 | -66.50 | -40.04 | -34.74 | 32.43 |
| Oregon | 1.56 | 66.11 | 31.14 | 7.03 | 49.24 | -149.03 | -62.51 | -88.64 | 10.64 | 70.92 |
| Pennsylvania | 63.86 | -333.95 | -0.19 | -100.5 | -18.24 | 89.23 | 0.35 | 74.65 | -0.53 | -14.73 |
| Rhode Island | 24.33 | 1.02 | 4.98 | -15.13 | -5.22 | 1.51 | -1.00 | -5.98 | -17.54 | -5.63 |
| South Carolina | 15.83 | 107.50 | -0.06 | 12.92 | -0.31 | -62.92 | -46.82 | 0.92 | 10.30 | 1.52 |
| South Dakota | 2.61 | 36.35 | -1.31 | -1.30 | 1.69 | -7.92 | 13.44 | -21.32 | 0.09 | 7.84 |
| Tennessee | 10.14 | 41.34 | -17.12 | 52.80 | -7.09 | -13.19 | 46.25 | -8.24 | -0.61 | -0.16 |
| Texas | -32.26 | 157.56 | -27.83 | -29.46 | -0.53 | -0.09 | -136.30 | 130.63 | 76.24 | 47.84 |
| Utah | 41.90 | 44.99 | -97.26 | -8.12 | 31.35 | -26.36 | 10.77 | 10.82 | 2.40 | 411.96 |
| Vermont | 1.74 | 14.43 | 3.65 | 1.48 | -0.23 | 0.00 | -24.32 | -56.09 | -11.38 | 4.64 |
| Virginia | -2.33 | -39.55 | 61.21 | -46.45 | 16.18 | -48.42 | -1.18 | 111.83 | 6.82 | 10.60 |
| Washington | 0.12 | 404.99 | 44.96 | -0.31 | 4.53 | -164.78 | -42.06 | -213.87 | 2.89 | 69.51 |
| West Virginia | 110.06 | -28.98 | -12.27 | 86.53 | 63.06 | -37.59 | 0.00 | 12.14 | -13.76 | -0.12 |
| Wisconsin | 95.47 | 2.33 | 5.32 | -5.20 | -2.93 | -15.30 | -6.89 | 0.36 | -14.03 | 0.47 |
| Wyoming | 32.76 | -18.45 | -3.58 | 14.86 | -4.52 | -1.14 | 1.34 | -4.48 | -5.05 | 58.59 |

**APPENDIX C Signed Chi-Square Tables**

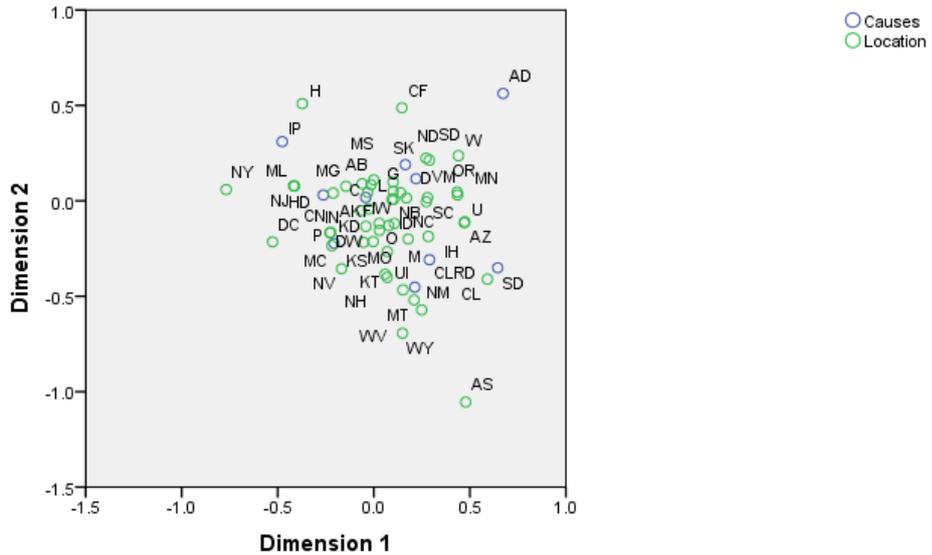

Row and Column Points

Symmetrical Normalization